\documentclass[12pt]{iopart}
\usepackage{graphicx}
\usepackage{dcolumn}
\usepackage{bm}

\def\aa{\rm aa}

\def\erf{{\rm erf}}      
\def\erfc{{\rm erfc}}

\def\no{\stackrel{o}{.}}

\def\erf{{\rm erf}}       
\def\erfc{{\rm erfc}}     

\def\calc{{\cal C}}

\def\aa{\rm aa}

\def\beq{\begin{equation}}
\def\ee{\end{equation}}
\def\eeq{\end{equation}}

\def\bfig{\begin{figure}}
\def\efig{\end{figure}}
\def\bea{\begin{eqnarray}}
\def\beann{\begin{eqnarray*}}
\def\eea{\end{eqnarray}}
\def\eeann{\end{eqnarray*}}

\def\nn{\nonumber}

\def\raw{\rightarrow}

\def\bi{\bibitem}

\begin{document}

\title{Do glueballs contribute in heavy ion collisions?}

\author{M. L. L. da Silva$^\dag$, D. Hadjimichef $^{\ddag\dag}$,  C. A. Z. Vasconcellos$^\dag$ and 
 B. E. J. Bodmann$^{\ast\,\ddag}$ }

\address{$^{\dag}$ Instituto de F\'{\i}sica, Universidade Federal do Rio Grande do  Sul,\\
 Av. Bento Gon\c{c}alves, 9500, Porto Alegre, R. S., CEP 91501-970, Brazil}

 \address{$^{\ast}$ Dep. de Engenharia Nuclear, Escola de Engenharia,\\
  Universidade Federal do Rio Grande do  Sul, Av. Osvaldo Aranha, 99 - 4$\no$ andar \\
  Porto Alegre, R. S., CEP 90035-190, Brazil}
 
\address{$^{\ddag}$ Instituto de F\'{\i}sica e Matem\'atica, Universidade Federal de  Pelotas,\\
 Campus Universit\'ario,  Pelotas, R. S., CEP 96010-900, Brazil}

\date{\today}

\begin{abstract}

In heavy ion collision simulations many hadron states and/or parton degrees of freedom are included in order to
obtain the observables. Meson spectroscopy, for example, considers the  $\,0^{++}$
meson as a mixture of $q\bar{q}$ and glue. This fact is usually not considered in heavy ion collision  physics.  In the present work we
consider two extreme possibilities for the
constitution of the $\,0^{++}$ meson,  either as a pure glueball or as $q\bar{q}$-meson. 
The scattering amplitude and cross-sections with constituent
interchange are determined for the two situations.
The comparison showed that the glueball-glueball elastic scattering
cross-section for a color singlet state 
is between one to two orders of magnitude smaller than the
corresponding $q\bar{q}\/$ state. The $\,2^{++}$ glueball-glueball
interaction is also evaluated with similar behavior.
Thus, glueball-glueball scattering is not very likely to introduce
significant changes in heavy ion collision observables.

\end{abstract}


\submitto{\JPG}

\maketitle

\section{Introduction}
\label{intro}

The existence of glueballs and their experimental detection remains a challenge, although proposed already three decades ago \cite{Fritsch75}.
Nevertheless, there are candidates in the scalar spectrum which have
not yet been uniquely identified and thus may be interpreted as
glueballs or hybrid states. The question of the existence of glueballs
is of particular interest for an understanding of heavy ion collision
signatures, because they may play a crucial role in the possible
formation of Quark Gluon Plasma (QGP). Such a phase transition may
involve the lightest possible glueball candidate with quantum numbers
$0^{++}\/$ and mass around $1.7$ GeV  \cite{Vento05}.
Some of the contributions to the phase transition are possibly states
known from hadron spectroscopy, however, it might also be necessary to
include color octet contributions. Spectroscopy so far supplies a less 
vague physical picture, hence we adopt a rather conservative point of
view and investigate the possible role of the $0^{++}\/$ (and
$2^{++}$) glueball state in heavy ion collision scenarios. The
present work is a first step in this direction, where we present a
model dependent description of two glueball states and their
interaction in order to investigate the question whether
glueball-glueball interaction is a candidate for a significant contribution in heavy ion reactions.

Although technical difficulties still trouble our understanding of
glueball properties in experiments, largely because glueball states
can mix strongly with near\-by  $q\bar{q}$ resonances
\cite{amsler1,amsler2}, recent experimental and lattice studies
of $0^{++}$, $2^{++}$ and $0^{-+}$ glueballs seem to be convergent
\cite{uthoma}-\cite{bali2}. 
On theoretical grounds, a simple
potential model with massive constituent gluons, namely the model of
Cornwall and Soni \cite{cs1,cs2} has received attention
recently \cite{cs3,cs4} for spectroscopic calculations.
In particular, one important issue, in this model, is the determination
of the lowest mass glueball: the $0^{++}$ resonance, which within the
model is  assumed to a have a mass $1.73$
GeV.  Experimentally  this resonance is now closely identified with the scalar
meson $f_0(1710)$ observed at Belle in $\gamma\gamma\raw K^{+}K^{-}(
K^{0}_{s}K^{0}_{s})$  \cite{belle} and 
 BES in $J/\psi\raw\gamma K^{+}K^{-}(\gamma K^{0}_{s}K^{0}_{s})$ \cite{bes}.
This resonance  is an isospin zero state so that in principal it
should be a mixture of the quark sector $q\bar{q}$ with a color singlet of glue  \cite{close}.
In particular there is growing evidence in the direction of large
$s\bar{s}$ content with 
some mixture with the glue sector. 

In the present work we consider two extreme possibilities for the
$0^{++}$ resonance: ({\it i}) as a glueball, calculating the
scattering amplitude and cross-section for a glueball-glueball
interaction, in the context of the constituent gluon model, with gluon
interchange; ({\it ii}) as a pure $s\bar{s}$ system, where a new
calculation is performed for the scattering amplitude and
cross-section for an $(s\bar{s})$-$(s\bar{s})$ interaction, in the
context of a quark interchange picture. However, it is not very likely
that exclusively either one or the other scenario is present in the
scattering process. Then for {\it s}-wave scattering the cross section
may be compared to experimental data by a Likelihood analysis, for
instance. To this end we treat glueball-glueball and
($s\bar{s}$)-($s\bar{s}$) interaction on the same theoretical footing:
In order to obtain a scattering amplitude and cross-section with constituent interchange, we follow the Fock-Tani formalism (FTf) approach \cite{annals}. This formalism is a mapping technique in which composite particles are mapped into ideal particles (with no sub-structure). The FTf shall be briefly described in the following section applied to glueballs, the case for a heavy ($q\bar{q}$)-($q\bar{q}$) system is described in detail elsewhere \cite{annals,sergio}.

\section{Fock-Tani Formalism for Glueballs}
\label{sec:1}

In the Fock-Tani representation one starts with the Fock representation of the system using field operators of elementary constituents which satisfy canonical (anti-)commu\-ta\-tion relations. Composite-particle field operators  are linear combinations of the elementary-particle operators and do not generally satisfy canonical (anti)commutation relations. ``Ideal" field operators acting on an enlarged Fock space are then introduced in close correspondence with the composite ones. The enlarged Fock space is a graded direct product of the original Fock space and an ``ideal state space". The ideal operators correspond to particles with the same quantum numbers of the composites; however, they satisfy by definition canonical (anti)commutation relations. Next, a given unitary transformation, which transforms the single composite states into single ideal states, is introduced. When the transformation acts on operators in the subspace of the enlarged Fock space which contains no ideal particles, the transformed operators explicitly express the interactions of composites and constituents. Application of the 
unitary operator on the microscopic Hamiltonian, or on other hermitian operators expressed in terms of the elementary constituent 
field operators, gives equivalent operators which contain the ideal field operators. The effective Hamiltonian in the new representation is hermitian and has a clear physical interpretation in terms of the processes it describes. Since all field 
operators in the new representation satisfy canonical (anti)commutation relations, the standard methods of quantum field theory can then be readily applied.

\begin{figure}
\begin{center}
\includegraphics[width=0.7\textwidth]{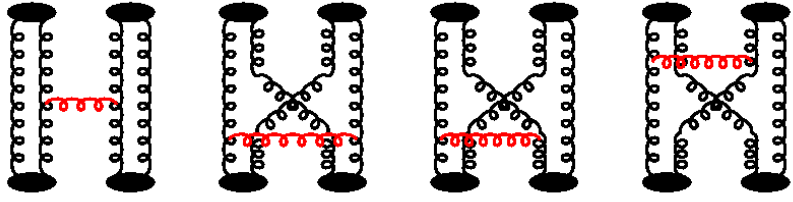}
\end{center}
\caption{Diagrams representing the scattering amplitude $h_{fi}$ for
  glueball-glueball interaction with constituent gluon interchange.}
\label{hfi_fig}
\end{figure}

The starting point in the present calculation is  the definition, in second quantization, of the glueball creation operator formed by two constituent gluons 
\bea
    G_{\alpha}^{\dagger}=\frac{1}{\sqrt{2}}\Phi_{\alpha}^{\mu \nu}
    a_{\mu}^{\dagger}a_{\nu}^{\dagger}.
\label{G}
\eea
Gluon creation $a^{\dag}_{\nu}$ and annihilation $a_{\mu}$ operators obey the conventional commutation relations
\bea
    [a_{\mu},a_{\nu}]=0 \,\,\,\, ; \,\,\,\,[a_{\mu},a_{\nu}^{\dagger}]=\delta_{\mu\nu}.
\eea
In (\ref{G}) $\,\Phi_{\alpha}^{\mu\nu}$ is the bound-state wave-function for two-gluons. The composite glueball operator satisfy non-canonical commutation relations
\bea
[G_{\alpha},G_{\beta}]=0\,\,\,\,\,;\,\,\,\, [G_{\alpha},G_{\beta}^{\dagger}]
    =\delta_{\alpha\beta}+\Delta_{\alpha\beta}
\eea
where
\bea
    \delta_{\alpha\beta}=\Phi_{\alpha}^{\star\rho \gamma}\Phi_{\beta}^{\gamma \rho}
\,\,\,\,\,\,;\,\,\,\,\,\,
    \Delta_{\alpha\beta}=2\Phi_{\alpha}^{\star\mu
      \gamma}\Phi_{\beta}^{\gamma \rho}a_{\rho}^{\dagger}a_{\mu}.
\eea
The  ``ideal particles'' which obey canonical relations, in our case  are the ideal glueballs
\bea
    [g_{\alpha},g_{\beta}]=0\,\,\,\,\,;\,\,\,\, [g_{\alpha},g_{\beta}^{\dagger}]
    =\delta_{\alpha\beta}.
\eea
This way one can transform the composite glueball state $|\alpha\rangle$  into an ideal state $|\alpha\,)$ by
\begin{eqnarray*}
 |\alpha\,)=   U^{-1}(-\frac{\pi}{2})\,G_{\alpha}^{\dagger}
\,|0\rangle=g_{\alpha}^{\dagger}\,| 0\rangle
\end{eqnarray*}
where $ U=\exp({tF})$ and $F$ is the generator of the glueball transformation given by
\begin{eqnarray}
    F=g_{\alpha}^{\dagger}\tilde{G}_{\alpha}-\tilde{G}_{\alpha}^{\dagger}g_{\alpha}
    \label{F}
\end{eqnarray}
with
\begin{eqnarray*}
    \tilde{G}_{\alpha}=G_{\alpha}-\frac{1}{2}\Delta_{\alpha\beta}G_{\beta}
    -\frac{1}{2}G_{\beta}^{\dagger}[\Delta_{\beta\gamma},G_{\alpha}]G_{\gamma}.
\end{eqnarray*}
In order to obtain the effective glueball-glueball potential one has to use (\ref{F}) in a set of Heisenberg-like equations for the basic operators $g,\tilde{G},a$
\begin{eqnarray*}
    \frac{dg_{\alpha}(t)}{dt}=[g_{\alpha},F]=\tilde{G}_{\alpha}\,\,\,\,\,;\,\,\,\,\,
    \frac{d\tilde{G}_{\alpha}(t)}{dt}=[\tilde{G}_{\alpha}(t),F]=-g_{\alpha}\,.
\end{eqnarray*}
The simplicity of these equations are not present in the equations for $a$
\begin{eqnarray*}
    \frac{da_{\mu}(t)}{dt}=[a_{\mu},F]=&-&\sqrt{2}\Phi_{\beta}^{\mu\nu}a_{\nu}^{\dagger}g_{\beta}
    +\frac{\sqrt{2}}{2}\Phi_{\beta}^{\mu\nu}a_{\nu}^{\dagger}\Delta_{\beta\alpha}g_{\beta}\\
    &+&\Phi_{\alpha}^{\star\mu\gamma}\Phi_{\beta}^{\gamma\mu^{'}}
    (G_{\beta}^{\dagger}a_{\mu^{'}}g_{\beta}-g_{\beta}^{\dagger}a_{\mu^{'}}G_{\beta})\\
    &-&\sqrt{2}(\Phi_{\alpha}^{\mu\rho^{'}}\Phi_{\rho}^{\mu^{'}\gamma^{'}}
    \Phi_{\gamma}^{\star\gamma^{'}\rho^{'}}\\
&&+\Phi_{\alpha}^{\mu^{'}\rho^{'}}\Phi_{\rho}^{\mu\gamma^{'}}
    \Phi_{\gamma}^{\star\gamma^{'}\rho^{'}})G_{\gamma}^{\dagger}a_{\mu^{'}}^{\dagger}
    G_{\beta}g_{\beta}.
\end{eqnarray*}
The solution for these equation can be found order by order in the wave-functions. For zero order one has $a_{\mu}^{(0)}=a_{\mu}$
\begin{eqnarray*}
    g_{\alpha}^{(0)}(t)&=&G_{\alpha}\sin{t}+g_{\alpha}\cos{t}\\
G_{\beta}^{(0)}(t) &=&G_{\beta}\cos{t}-g_{\beta}\sin{t}.
\end{eqnarray*}
In the first order $g_{\alpha}^{(1)}=0,\,\,\,G_{\beta}^{(1)}=0$ and
\begin{eqnarray*}
    a_{\mu}^{(1)}(t)=\sqrt{2}\Phi_{\beta}^{\mu\nu}a_{\nu}^{\dagger}[G_{\beta}^{(0)}-G_{\beta}].
\end{eqnarray*}
The second order expression is
\begin{eqnarray*}
    a_{\mu}^{(2)}(t)&=&-2\Phi_{\alpha}^{\star\mu\gamma}\Phi_{\beta}^{\gamma\mu^{'}}
    G_{\beta}^{\dagger}a_{\mu^{'}}G_{\alpha}^{(0)}
+\Phi_{\alpha}^{\star\mu\gamma}    \Phi_{\beta}^{\gamma\mu^{'}}G_{\beta}^{\dagger}a_{\mu^{'}}G_{\alpha}
\nonumber\\
&&+\Phi_{\alpha}^{\star\mu\gamma}
\Phi_{\beta}^{\gamma\mu^{'}}G_{\beta}^{\dagger(0)} a_{\mu^{'}}G_{\alpha}^{(0)}.
\end{eqnarray*}
To obtain the third order $a_{\mu}^{(3)}(t)$ is straightforward
\begin{eqnarray*}
    a_{\mu}^{(3)}(t)&=&\sqrt{2}\Phi_{\alpha}^{\mu\nu}\Phi_{\beta}^{\star\sigma\nu}
    \Phi_{\gamma}^{\sigma\tau}(G_{\beta}^{\dagger}a_{\tau}^{\dagger}G_{\gamma}G_{\alpha}^{(0)}
    -G_{\beta}^{\dagger(0)}a_{\tau}^{\dagger}G_{\gamma}G_{\alpha}^{(0)}\nonumber\\
    &+&G_{\beta}^{\dagger(0)}a_{\tau}^{\dagger}G_{\gamma}^{(0)}G_{\alpha}^{(0)}
    -G_{\beta}^{\dagger}a_{\tau}^{\dagger}G_{\gamma}G_{\alpha})\nonumber\\
    &-&\frac{\sqrt{2}}{2}\Phi_{\alpha}^{\mu\nu}a_{\nu}^{\dagger}\Delta_{\alpha\gamma}
    [(\cos{t}-2)G_{\gamma}+G_{\gamma}^{(0)}].
\end{eqnarray*}
It is sufficient to evaluate up to third order, because the effective interaction is forth ordered in the transformed operators.
The glueball-glueball potential can be obtained applying in a standard way the Fock-Tani transformed operators to the microscopic Hamiltonian
\begin{eqnarray*}
    {\cal{H}}(\mu\nu;\sigma\rho)=T_{\rm aa}(\mu)a_{\mu}^{\dagger}a_{\mu}+\frac{1}{2}
    V_{\rm aa}(\mu\nu;\sigma\rho)a_{\mu}^{\dagger}a_{\nu}^{\dagger}a_{\rho}a_{\sigma}
\end{eqnarray*}
where in this microscopic Hamiltonian $T_{\rm aa}$ is the kinetic energy and $V_{\rm aa}$ is the  potential in the constituent model. After transforming  $ {\cal{H}}(\mu\nu;\sigma\rho)$ one obtains for the glueball-glueball potential $V_{gg}$
\bea
    V_{gg}=\sum_{i=1}^{4}V_{i}(\alpha\gamma;\delta\beta)g_{\alpha}^{\dagger}
    g_{\gamma}^{\dagger}g_{\delta}g_{\beta}
\label{v_gg}
\eea
and
\bea
    &V_{1}(\alpha\gamma;\delta\beta)=2V_{aa}(\mu\nu;\sigma\rho)\Phi_{\alpha}^{\star\mu\tau}
    \Phi_{\gamma}^{\star\nu\xi}\Phi_{\delta}^{\rho\xi}\Phi_{\beta}^{\sigma\tau}\nn\\
    &V_{2}(\alpha\gamma;\delta\beta)=2V_{aa}(\mu\nu;\sigma\rho)\Phi_{\alpha}^{\star\mu\tau}
    \Phi_{\gamma}^{\star\nu\xi}\Phi_{\delta}^{\rho\tau}\Phi_{\beta}^{\sigma\xi}\nn\\
    &V_{3}(\alpha\gamma;\delta\beta)=V_{aa}(\mu\nu;\sigma\rho)\Phi_{\alpha}^{\star\mu\nu}
    \Phi_{\gamma}^{\star\lambda\xi}\Phi_{\delta}^{\sigma\lambda}\Phi_{\beta}^{\rho\xi}\nn\\
    &V_{4}(\alpha\gamma;\delta\beta)=V_{aa}(\mu\nu;\sigma\rho)\Phi_{\alpha}^{\star\mu\xi}
    \Phi_{\gamma}^{\star\nu\lambda}\Phi_{\delta}^{\lambda\xi}\Phi_{\beta}^{\rho\sigma}\,.
\label{v1-v4}
\eea
The scattering $T$-matrix is related directly to equation (\ref{v_gg}) 
\bea
T(\alpha\beta;\gamma\delta)=(\alpha\beta|V_{gg}|\gamma\delta)\,.
\label{t-matrix}
\eea
Due to translational invariance, the $T$-matrix element is written as a momentum conservation delta-function, times a Born-order matrix element, $h_{fi}$:
\bea
T(\alpha\beta;\gamma\delta)=\delta^{(3)}( \vec{P}_{f}-\vec{P}_{i})\,h_{fi}
\label{t-hfi}
\eea
where $\vec{P}_{f}$ and  $\vec{P}_{i}$ are the final and initial momenta of the two-glueball system.  This result can be used
in order to evaluate  the glueball-glueball scattering cross-section 
\bea
\sigma_{gg} =\frac{4\pi ^{5}\,s}{s-4M_{G}^{2}}
\int_{-(s-4M_{G}^{2})}^{0}\,dt\,|h_{fi}|^{2}
\label{cross}
\eea
where $M_G$ is the glueball mass, $s$ and $t$ are the Mandelstam variables. The scattering amplitude $h_{fi}$ can be visualized in figure \ref{hfi_fig}.

\section{The Constituent Gluon Model}

\subsection{The mass equation}

In connection with the general formalism presented in the last section, we now concentrate in a particular microscopic model.   The potential $V_{\rm aa}$ is chosen in the context of the Cornwall and Soni constituent gluon model (CGM) \cite{cs1}. Their original work extended to study two-gluon and three-gluon glueballs. Gluon dynamics was described as massive spin-one fields 
interacting through massive spin-one exchange and by a breakable string. In the present we  shall restrict our calculation to the two-gluon sector and total $L=0$. The CGM Hamiltonian is given by 
\bea
  H_{\aa} = 2m -\frac{1}{m}\nabla^2 + V_{\aa}
  \label{hamilt}
\eea
where $m$ is the effective gluon mass and
\bea
V_{\rm aa}(r)= \frac{1}{3}\, f^{ace}f^{bde}   \left[\,V^{\rm OGEP}(r)+V_{S}(r)\right].
\label{mod24}
\eea
The potential $V_{\rm aa}$  in (\ref{mod24}) contains a one-gluon exchange term $V^{\rm OGEP}$
\bea
V^{\rm OGEP}(r)&=&   -\alpha\,
 \left[ \,\omega_{1}\frac{e^{-m r}}{r}
+\omega_{2}\frac{\pi}{m^2}\,F(r)\,
\right]
\label{mod24a}
\eea
where  $\alpha = 3\,\alpha_{s}$, $\omega_{1}$ and $\omega_{2}$ are defined as 
\bea
\omega_{1}& = &\frac{1}{4}+\frac{1}{3}\vec{S}^2 
\,\,\,\,\,\,\,\,\,\,,\,\,\,\,\,\,\,\,\,\,
\omega_{2} = 1-\frac{5}{6}\vec{S}^2  
\label{omega}
\eea
with $\vec{S}$ is the glueball's total spin. In (\ref{mod24a}) there appears a form factor $F(r)$, which beside the explicitly considered interactions parametrizes a residual interaction. This additional contribution is necessary to avoid a collapse of the glueball $0^{++}\/$ state and is set to be a Gaussian type with parameter $\rho$
\bea
F(r)&=&\frac{\rho^{3}}{\pi^{3/2}}\,e^{-\rho^{2}r^{2}}.
\label{mod24b}
\eea
The second term in (\ref{mod24}) is the string potential responsible for the confinement 
\bea
V_{S}(r)&=& 2m\,(1-e^{-\beta \,m\,r})
\label{mod24c}
\eea
where $\beta$ is the string tension.

The scattering amplitude is central in the calculation and shall depend on the parameters that appear in $V_{\rm aa}$, namely $\alpha$, $\beta$, $m$, $\rho$. A criterion for adjusting these parameters is that the expectation value of $H_{\rm aa}$ should render the glueball's mass
\bea
\Phi_{\alpha}^{\ast\mu\nu}  H_{\aa}(\mu\nu;\sigma\rho)
 \Phi_{\beta}^{\sigma\rho} = M_G.
\label{valor-es}
\eea
The glueball's wave-function $\Phi$ is written as a product
\bea
\Phi_{\alpha}^{\mu\nu}=
\chi_{A_{\alpha}}^{s_{\mu}s_{\nu}}\,
{\cal C}^{c_{\mu}c_{\nu}}\,
\Phi_{\vec{P}_{\alpha} }^{\,\vec{p}_{\mu}\vec{p}_{\nu}},
\label{wf}
\eea
$\chi_{A_{\alpha}}^{s_{\mu}s_{\nu}}$ is the spin contribution, with $A_{\alpha}\equiv \{S_{\alpha},S^{3}_{\alpha}\} $, where $S_{\alpha}$ is the glueball's  total spin index  and $S^{3}_{\alpha} $ the index of the spin's third component; ${\cal C}^{c_{\mu}c_{\nu}}$ is the color component given by
\bea
{\cal C}^{c_{\mu}c_{\nu}}=\frac{1}{\sqrt{8}}
\,\delta^{c_{\mu}c_{\nu}}
\label{phi-cor}
\eea
and the spatial wave-function is
\bea
\Phi_{\vec{P}_{\alpha}}^{\vec{p}_{\mu}\vec{p}_{\nu}}=
\delta^{(3)}(\vec{P}_{\alpha}-\vec{p}_{\mu}-\vec{p}_{\nu})\,
\phi(\vec{p}_{\mu}-\vec{p}_{\nu})\;,
\label{phi-7}
\eea
where
\bea
\phi(\vec{p}_{\mu}-\vec{p}_{\nu})={\left(\frac{1}{\pi
b^2}\right)}^{\frac{3}{4}}e^{-\frac{1}{8b^{2}}{
\left(\vec{p}_{\mu}-\vec{p}_{\nu}\right)}^{2}}.
\label{phi-8}
\eea
The expectation value of $r^2$ gives a relation between the $rms$ radius $r_0$ and $b$
\bea
b=\frac{\sqrt{1.5}}{r_0}.
\label{b-r0}
\eea
These relations when introduced in (\ref{valor-es}) result in the {\it mass equation} which relates all parameters in the model
\bea
M_{G}&=&4m +\frac{3b^2}{2m}
-  \frac{\omega_1 \alpha}{\sqrt{\pi} } \left[2b + m \,\sqrt{\pi}
  \, e^{\frac{m^2}{4b^2}}\, \erfc{\left(\frac{m}{2b}
  \right)} \right]
\nn\\
&&
-
  \frac{m \zeta^3 \omega_2 \alpha }{\sqrt{\pi}}\,\,\,
  \frac{b^3}{(b^2 + \zeta^2 m^2)^{3/2}}
+
 \frac{m}{\sqrt{\pi}b^2} \left[ 2b \beta m 
\frac{}{}\right.
\nn\\
&&
\left.
+ \sqrt{\pi}
  \,e^{\frac{\beta^2 m^2}{4b^2}} (2 b^2 + \beta^2 m^2)
\,  \erfc{\left(\frac{\beta m}{2 b} \right)} \right]
\label{mg}
\eea
where $\zeta=\rho/m$. The complementary error function is $\erfc(x) \equiv \erf(x) -1$ with
\bea
  \erf(x) = \frac{2}{\sqrt{\pi}} \int_{0}^{x} dt \,\,e^{-t^2}.
  \label{erf}
\eea
To adjust the parameters $\rho$ and $\beta$ without ambiguity we shall use the mass equation not only for $0^{++}$ but also for the next low-lying glueball candidate, in the $\vec{L}=0$ sector, the $2^{++}$ resonance. The results obtained from lattice QCD, for these resonances give a mass estimate for $M_{0^{++}}\approx 1.7$ GeV and $M_{2^{++}}\approx 2.4$ GeV. An additional result from lattice is that independent from the absolute mass values, the mass ratio shall be
\bea
  \frac{M_{2^{++}}}{M_{0^{++}}} \simeq 1.4.
  \label{razao}
\eea


The model parameters were fixed using a parametric inference method
reported in detail elsewhere \cite{SilvaHB06}, in order to calibrate
the model for subsequent cross-section calculations.
To this end we used a Likelihood Monte Carlo method that
simultaneously fixed the parameters for 
the $0^{++}\/$ and the $2^{++}\/$ candidates. The following
restrictions for the parameters were implemented: The string tension
$\beta\/$ shall be common for both states since they shall appear in
the same spectrum. The parameter $b\/$ for either state shall be of
comparable magnitude, also the form-factor parameter $\zeta\/$ shall
be comparable. The effective gluon mass shall be roughly an order of
magnitude smaller than the glueball state \cite{cs3,cs4}. The
glueball masses $M_G\/$ for the quantum numbers $0^{++}\/$ and
$2^{++}\/$ shall be in the expected mass ranges given by Lattice calculations.

Maximizing the combined likelihood function for both states yielded the best parameter sets for both sates simultaneously. All parameters except for the glueball masses were generated randomly within a given interval and using an initially homogeneous distribution. The parameter combinations were accepted in a two step evaluation, first their glueball masses were to fit in an interval with a width $10 \%\/$ of the respective glueball mass around the value determined from lattice QCD calculations. In a subsequent step, parameter combinations were excluded where the radius values $b(0^{++})\/$ and $b(2^{++})\/$ differed by more than a factor of two.
From $10^5\/$ generated parameter combinations approximately half the number were accepted by the Likelihood criterion, fitting the two states in the same spectrum. The simultaneous glueball mass selection reduced the set to $14\/$ candidates from which three survived the elimination from the $b\/$ and $\zeta\/$ value differences of the respective $0^{++}\/$ and $2^{++}\/$ glueball states. The table  \ref{parset} shows the three accepted parameter sets.

The gluon coupling constant $\alpha$ is three times larger than the
strong coupling constant $\alpha_s$. The usual values for $\alpha_s$
is the order of $0.6$, which sets $\alpha$ to $1.8$. From the $rms$
radius $r_0$ relation to the wave-function's $b$ parameter,
Eq. (\ref{b-r0}), the usual hadronic range is reproduced with $r_0 \approx 0.1 - 0.8 fm\/$.

\begin{table}
\caption{Parameter sets for glueball candidates with $m\/$, $b\/$ and $M\/$ in units of $GeV\/$.}
\begin{center}
\begin{tabular}{lcccccccc}
\hline
Set & $\beta$ & $m$ & $b(0^{++})$ & $b(2^{++})$ & $\zeta(0^{++})$ & $\zeta(2^{++})$ & $M(0^{++})$ & $M(2^{++})$ \\
\hline\hline
(a) & $1.7$ & $0.1$ & $0.4$ & $0.5$ & $0.2$ & $0.2$ & $1.9$ & $2.3$ \\
(b) & $5.8$ & $0.1$ & $0.3$ & $0.5$ & $0.2$ & $0.2$ & $2.0$ & $2.4$ \\
(c) & $2.2$ & $0.1$ & $0.4$ & $0.6$ & $0.2$ & $0.2$ & $1.8$ & $2.4$ \\
\hline
\end{tabular}
\end{center}
\label{parset}
\end{table}


\subsection{The scattering amplitude and cross-section}

\begin{figure}
\begin{center}
\includegraphics[width=0.4\textwidth,angle=-90]{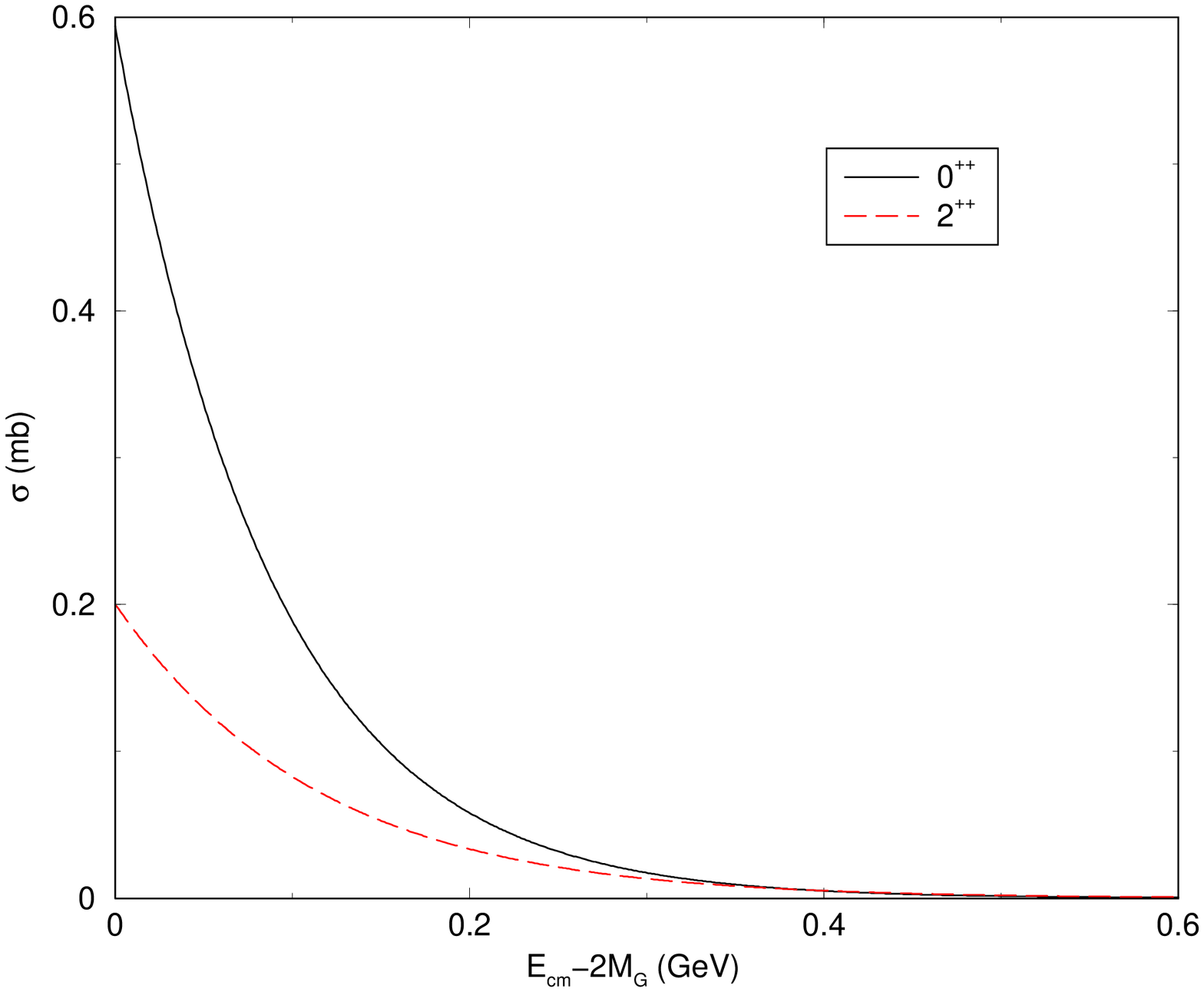}
\includegraphics[width=0.4\textwidth,angle=-90]{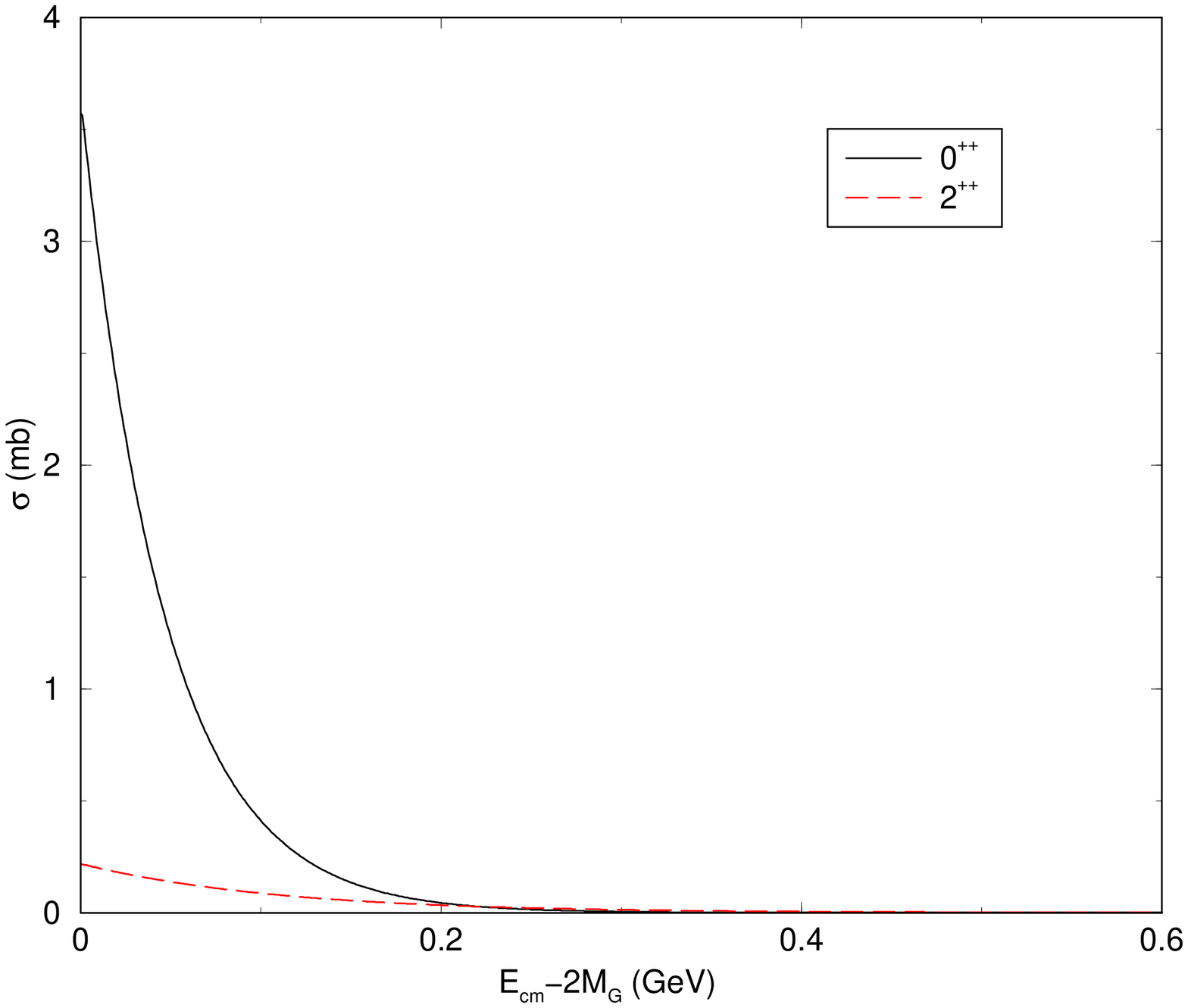}
\includegraphics[width=0.4\textwidth,angle=-90]{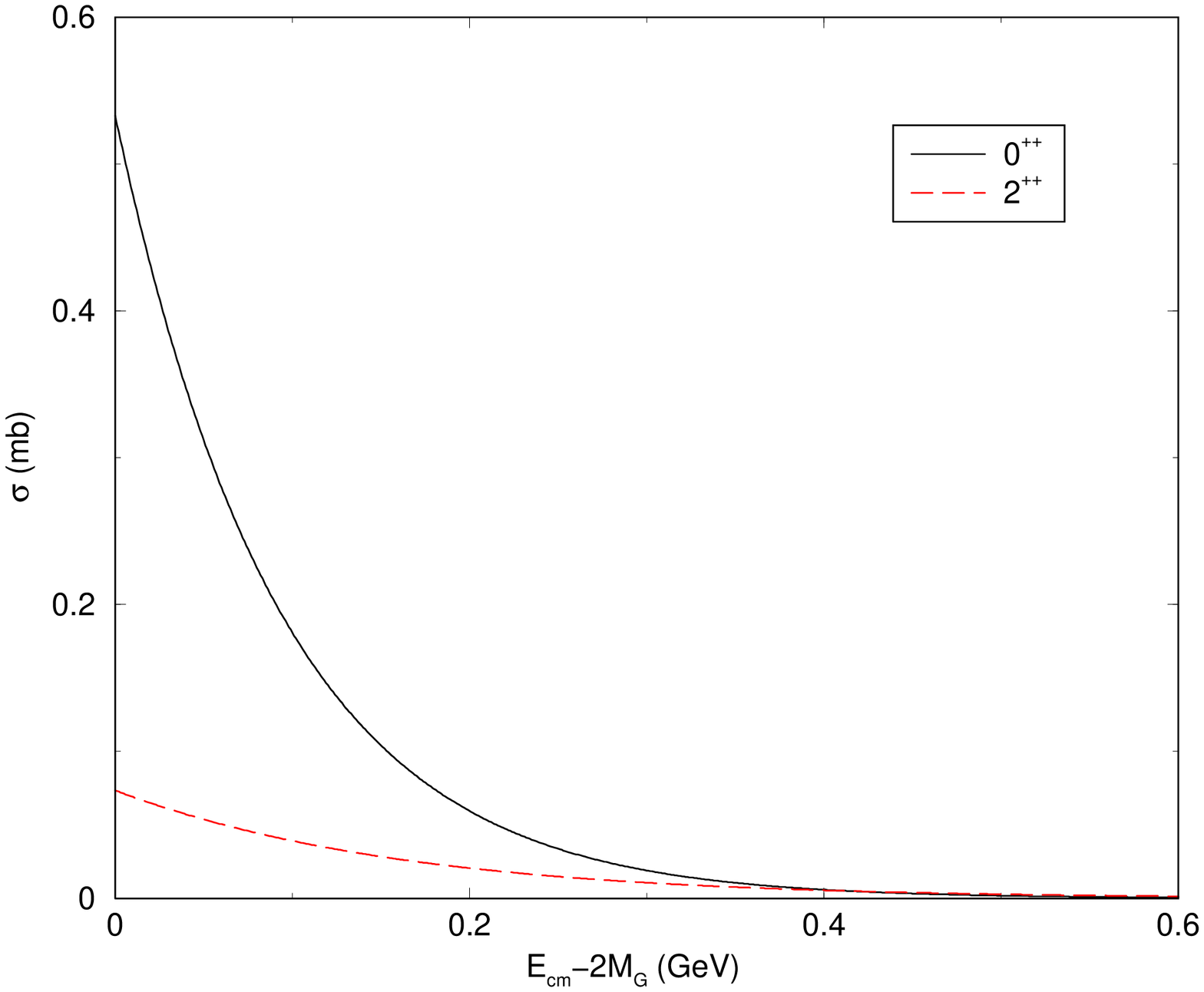}
\end{center}
\caption{Elastic cross-section for three parameter sets, upper left
  $\beta=1.7\/$, $b(0^{++})=0.4\/$, $b(2^{++})=0.5\/$,
  $M(0^{++})=1.9\/$, $M(2^{++})=2.3\/$; upper right $\beta=5.8\/$,
  $b(0^{++})=0.3\/$, $b(2^{++})=0.5\/$, $M(0^{++})=2.0\/$,
  $M(2^{++})=2.4\/$ and lower $\beta=2.2\/$, $b(0^{++})=0.4\/$,
  $b(2^{++})=0.6\/$, $M(0^{++})=1.8\/$, $M(2^{++})=2.4\/$.}
\label{csg}
\end{figure}

In order to evaluate the cross-section given by equation (\ref{cross})
one has to obtain the scattering amplitude $h_{fi}$ from  diagrams of figure
\ref{hfi_fig}. These diagrams  are explicitly calculated as 
multidimensional integrals in the momentum space from Eqs. 
(\ref{v1-v4})-(\ref{t-hfi}). The choice of a Gaussian wave-function for
the spatial part of $\Phi$ in (\ref{phi-8}), provides a simplification in the integration 
and retains the basic ingredients of color confinement. Schematically the evaluation of
the scattering amplitude can be written as the following  product
\bea
h_{fi}(s,t)=\sum_{i} {\cal C}_{i}\,h_{i}(s,t,\omega_{1}^{(i)},\omega_{2}^{(i)} )
\eea
 Details of the spatial, color and spin factors calculation are presented in the appendix.
Here we present the final result
\bea
  h_{fi}(s,t) &=& \frac{3}{8}\,R_{0}(s)
\sum_{i=1}^{6}\,R_{i}(s,t)
\label{hfi-st}
\eea
where
\bea
R_{0}&=& \frac{4}{(2\pi)^{3/2}b^3}
  \exp\left [-\frac{1}{2b^2} \left (\frac{s}{4} -M_{G}^2 \right)  \right]\nn\\
R_{1}&=&\frac{\alpha\, \omega_{1}^{(2)}\, 4\,\sqrt{2\pi}}{3     }  \,\,
  \int_{0}^{\infty} dq\,\frac{q^2}{q^2 +m^{2}}
  \exp\left(-\frac{q^2}{2b^2}\right) \nn\\
&&\times\left[
{\mathcal{J}}_0  \left(\frac{q\sqrt{t}}{2b^2}\right) +
{\mathcal{J}}_0  \left(\frac{q\sqrt{u}}{2b^2}\right)
\right] 
\nn\\
R_{2}&=&
\frac{  \alpha\, \omega_{2}^{(2)}     2\sqrt{2} \pi b^3 \zeta^3 m}{3(b^2
  +2\zeta^2 m^{2})^{3/2}} 
\left[
\exp \left( -\frac{t \zeta^2 m^2}{4(b^4 +2b^2 \zeta^2 m^{2})}\right)
\right.
\nn\\
&&
+\left.
\exp \left( -\frac{u \zeta^2 m^2}{4(b^4 +2b^2 \zeta^2 m^{2})}\right)
\right]\nn\\
R_{3} &=&
\frac{32\sqrt{2\pi}}{3} \int_0^{\infty} dq\,
  \frac{q^2 \beta  m^{2}}{(q^{2} +\beta^{2}m^{2})^{2}} \exp
  \left(-\frac{q^2}{2b^2}\right)
\nn\\
&&\times\left[
 {\mathcal{J}}_0 \left(\frac{q\sqrt{t}}{2b^2}  \right)+
 {\mathcal{J}}_0 \left(\frac{q\sqrt{u}}{2b^2}  \right)
\right]\nn\\
R_{4}&=&
 - \frac{16  \alpha\, \omega_{1}^{(3)} \sqrt{2\pi}\,b^2}{3\sqrt{\frac{s}{4}-M_{G}^2}}
 \int_{0}^{\infty} dq\,\frac{q}{q^2 +m^{2}}
\nn\\
 &&\times \exp\left(-\frac{3q^2}{8b^2}\right)
  \sinh\left(\frac{q}{2b^2}\sqrt{\frac{s}{4}-M_{G}^2}\, \right) \nn\\
R_{5}&=&
-\frac{16\alpha\, \omega_{2}^{(3)} \pi b^3 \zeta^3 m}{3(2b^2
  +3\zeta^2 m^{2})^{3/2}} \exp \left[ -\frac{\zeta^2 m^2 \left( \frac{s}{4}
  -M_{G}^2\right)}{2(2b^4 +3b^2 \zeta^2 m^{2})}\right]\nn\\
R_{6}&=&-\frac{128 \sqrt{2\pi}
  b^2}{3\sqrt{\frac{s}{4} -M_{G}^2}} \int_0^{\infty} dq\,
  \frac{q \beta  m^{2}}{(q^{2} +\beta^{2}m^{2})^{2}}
\nn\\
 &&\times \exp \left(\frac{3q^2}{8b^2}\right)  \sinh \left(\frac{q}{2b^2}
  \sqrt{\frac{s}{4}-M_{G}^2}\, \right).
\label{R}
\eea
In  (\ref{R}), ${\cal J}_{0} $ represents the spherical Bessel function defined by
${\cal J}_{0}(x)=\sin x /x $. The notation $\omega^{(i)}_{1}$
and $\omega^{(i)}_{2}$ is introduced, where the index $i$ corresponds to the number of the evaluated 
diagram in figure \ref{hfi_fig}. The cross-section for  scattering of the glueballs with
constituent gluon interchange  is obtained numerically by  inserting Eqs. 
(\ref{hfi-st}) and (\ref{R}) in (\ref{cross}). Assigning values as determined from parametric inference to $M_{G}$ in order
to correspond to the glueball $0^{++}$ ($M_{G} \approx  1.8-2.0$ GeV) and $2^{++}$ ($M_{G}\approx 2.3-2.4$ GeV)
one obtains the energy dependence of the elastic scattering cross-sections (see graphs in figure \ref{csg}).

\section{The Constituent Quark Model}

In order to discuss meson-meson scattering with constituent quark interchange, 
one needs to specify  the general form of the microscopic quark Hamiltonian. For our purposes here,
the microscopic Hamiltonian can be written in terms of the quark and antiquark
operators as 
\begin{eqnarray}
H&=& T(\mu)\; q_{\mu}^{\dagger }q_{\mu}+T(\mu)\; \bar{q}_{\mu}^{\dagger }\bar{q}_{\mu}
 +\frac{1}{2} V_{qq}(\mu \nu; \sigma\rho) \; q_{\mu}
^{\dagger }q_{\nu}^{\dagger }q_{\rho} q_{\sigma}\nn\\
&+&  \frac{1}{2} V_{\bar{q}\bar{q}}(\mu \nu; \sigma\rho)
\bar{q}_{\mu}^{\dagger }\bar{q}_{\nu}^{\dagger }\bar{q}_{\rho}\bar{q}_{\sigma}
+ V_{q\bar{q}}(\mu \nu; \sigma\rho)  
q_{\mu}^{\dagger }\bar{q}_{\nu}^{\dagger }\bar{q}_{\rho} q_{\sigma}
\label{ft23}
\end{eqnarray}
where $T$ is the kinetic energy; $V_{qq}$, $V_{\bar{q}\bar{q}}$,
$V_{q\bar{q}}$ are the quark-quark, antiquark-antiquark and quark-antiquark interactions.
In our calculation we shall use, for $V_{qq}$ , the 
spin-spin hyperfine component of the perturbative one gluon interaction
\bea
V_{qq} =  -\,\frac{8\pi\alpha_{s}}{3\,m_{i}m_{j}} \,
 {\bf S}_{i}\cdot{\bf S}_{j} \,\,\, {\cal F}^{\,a}_{i}\,\, {\cal F}^{\,a}_{j}
 \,\,\,,
\label{vss}
\eea
where ${\cal F}^{\,a}_{i}=\lambda^{a}_{i}/2$ are the Gell-Mann matrices. To obtain  $V_{\bar{q}\bar{q}}$,
one substitutes $\lambda \raw -\lambda^{T} $ and for  $V_{q\bar{q}}$ the following  
$\lambda_{j} \raw -\lambda^{T}_{j} $ in (\ref{vss}).

There is a considerable 
literature related to free meson-meson and baryon-baryon
scattering with constituent interchange \cite{oka1} - \cite{dimi9}. In many of these models the 
potential is much more elaborated than  (\ref{vss}) 
(including Coulomb, spin-orbit, tensor, confinement terms and eventually meson coupling to quarks). 
The lesson taken from all of these approaches is that the dominant term for the short-range repulsion is 
basically the spin-spin term from the  one gluon exchange potential. Its strong influence is seen, 
for example, in the $^{1}S_{\,0}$ partial-wave.

\begin{figure}
\begin{center}
\includegraphics[width=0.6\textwidth]{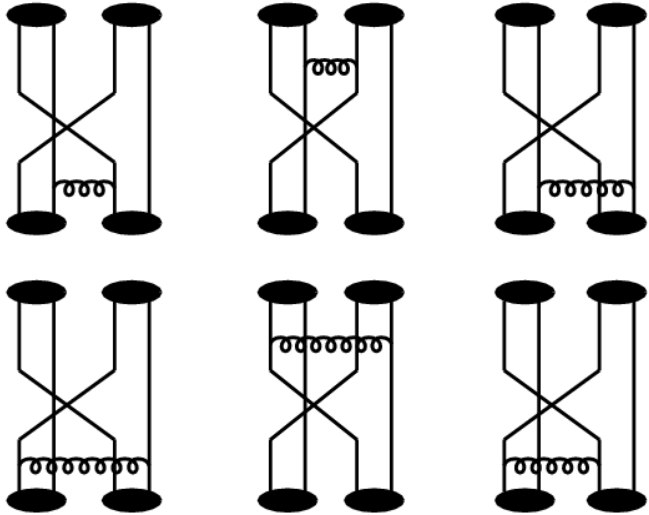}
\end{center}
\caption{Diagrams representing the scattering amplitude $h_{fi}$ for
  meson-meson interaction with constituent quark interchange.}
\label{meson_fig}
\end{figure}

The Fock-Tani formalism has been used in this context to study heavy meson scattering by pions
as described in reference \cite{annals} and \cite{sergio}. The scattering amplitude obtained
is  represented in figure \ref{meson_fig}.
The corresponding cross-section
for a $0^{++}$ meson with a $s\bar{s}$ content can be derived directly from this previous result by
a simple substitutions: the quark's mass $m_{q}\raw m_{s}$ and
the meson's mass by the glueballs $M\raw M_G\/$. The
remaining coefficients are unchanged. As described in \cite{annals} the meson wave-function is Gaussian
which implies exact analytical results for the scattering amplitude 
\bea
h_{fi}^{(s\bar{s})-(s\bar{s})}&=&
\frac{ 8\pi\alpha_s}{ 9m^{2}_{s}    (2\pi)^3    }
\left[ \frac{16  }{3\sqrt{3}} e^{-\frac{\xi}{12b^2}} 
+e^{\frac{t}{8b^2}} 
+e^{-\frac{u}{8b^2}}
\right]\nn\\
\label{hfi-ss}
\eea
with $\xi=s-4M_{G}^2$. The parameter $b$ has an equivalent origin as for the glueball, it is
the Gaussian length parameter and is related to the meson's {\it rms} radius $r_0$ by the same
expression as before: Eq.  (\ref{b-r0}).
The cross-section is obtained inserting Eq. (\ref{hfi-ss}) 
in (\ref{cross}) which results in
\bea
\sigma^{(s\bar{s})-(s\bar{s})}&=&\frac{4\pi\alpha_{s}^{2}s}{81m_{s}^{4}}
\left[
\frac{4b^2\left(1-e^{-\frac{\xi}{4b^2}}
\right)}{\xi}+\frac{128}{27}e^{-\frac{\xi}{6b^2}}
\right.
\nn\\
&&
\left.
+e^{-\frac{\xi}{8b^2}}
+\frac{64}{3\sqrt{3}}\frac{4b^2}{\xi}\left(e^{-\frac{\xi}{12b^2}}
-e^{-\frac{5\xi}{24b^2}}
\right)\right]\;\nonumber
\eea

\begin{figure}
\begin{center}
\includegraphics[width=0.4\textwidth,angle=-90]{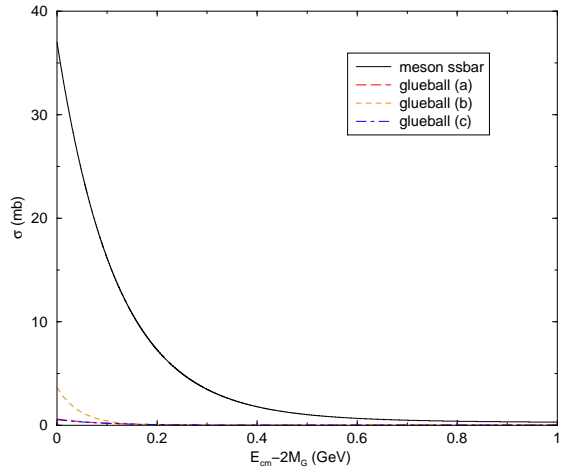}
\end{center}
\caption{Cross-section comparison for $0^{++}$ states.}
\label{cross0mm}
\end{figure}

The comparison between the cross-sections in the glueball picture and the 
quark picture for the $0^{++}$ meson is given in figure \ref{cross0mm}.
There is a sensitive difference in the cross-sections as a function of their
internal structure. For low energies it is  a notable that the quark-$0^{++}$
reaches high vales, while the gluon-$0^{++}$ approaches zero.
For higher CM  energies the curves have a similar behavior, even though
the $rms$ have  different values. 
The same comparison for the $2^{++}$ is beyond the scope of the present work, which
would imply in contributions with $\vec{L}\neq 0$ in order to build
the correct angular momentum states.

\section{Conclusions}

In the present article we compared two elastic scattering scenarios for the $0^{++}\/$ meson with mass $\sim 1.7$ GeV: one in which it was considered as a pure glueball and another as a $s\bar{s}\/$ state.
No particular mixing scheme was introduced. A  scattering cross-section for the $2^{++}\/$ $s\bar{s}$-meson,  with mass $M \sim 2.4$ GeV, was also calculated. The same comparison for this case was beyond the scope of the present work, which would imply in contributions with $\vec{L}\neq 0$ in the CGM in order to build
the correct angular momentum states.

The comparison, for the  $0^{++}\/$ meson, showed that the glueball-glueball elastic scattering cross-section for a color singlet state from 
is between one to two orders of magnitude smaller than the corresponding $q\bar{q}\/$ 
state. 
The glueball-glueball cross-section for  $2^{++}\/$  presented the same behavior.
Thus, if added as a new channel to already existing modes in presently used transport codes like the UrQMD,
glueball-glueball scattering is not very likely to introduce significant changes in, for example, the flow properties. Though, glue states whether pure or mixed are believed to somehow play a manifest role in collisions. However, there still might be a contribution from $q\bar{q}\/$-glue mixed states, once defined the particular
mixing scheme, with correspondence in hadronic spectra. This effect which we shall investigate in a subsequent work.

If our result holds for other color singlet states containing pure
glue, then it seems that states identified in hadron spectroscopy,
with stronger evidence as glueball or glue rich states, do not
contribute significantly to signatures in heavy ion reactions and one
shall resort to new possibilities as already indicated by lattice
calculations \cite{mstar,bali}, meaning that hadronic states known
inside a hadronic environment with high density and temperature have
no or almost no vacuum counter parts. Results from lattice
calculations \cite{mstar,bali} point towards a direction, where a
number of new states shall be considered and tested whether they suit
to explain the apparently low viscose flow property, which is
considered a strong indication for QGP. At high temperatures bulk
states of quarks, gluons and mixed states may exist and form
quasi-particles \cite{Hansen98}, which might explain the almost perfect liquid property of the plasma.

\vspace{1cm}

{\bf Acknowledgments }

\vspace{.3cm}

The authors would like to thank H. St\"ocker, J. Aichelin and W. Greiner
for important and enlightening discussions at 2$^{\rm \,nd}$ International Workshop on
Astronomy and Relativistic Astrophysics (IWARA) held at
Natal, R.N., Brazil, 2005.

\vspace{1cm}

\appendix

\section{Evaluation of the color factor, spin matrices and spatial integrals 
for glueball-glueball interaction}

The color  factor  ${\cal C}_{i}$, is calculated for each  diagram of
figure  \ref{hfi_fig},
\bea
  {\cal C}_{1} & = & f^{\mu\sigma\lambda}
  f^{\nu\rho\lambda}\,\, 
\calc^{\mu\tau}\calc^{\nu\xi}
  \calc^{\rho\xi}\calc^{\sigma\tau}\nn\\
  {\cal C}_{2} & = & f^{\mu\sigma\lambda}  f^{\nu\rho\lambda}\,\, 
\calc^{\mu\tau}\calc^{\nu\xi}
  \calc^{\rho\tau}\calc^{\sigma\xi}\nn\\
  {\cal C}_{3} & = &f^{\mu\sigma\lambda}
  f^{\nu\rho\lambda}\,\, \calc^{\mu\nu}\calc^{\tau\xi}
  \calc^{\sigma\tau}\calc^{\rho\xi}\nn\\
  {\cal C}_{4} & = & f^{\mu\sigma\lambda}
  f^{\nu\rho\lambda}\,\, \calc^{\mu\xi}\calc^{\nu\tau}
  \calc^{\tau\xi}\calc^{\rho\sigma}.
\label{color}
\eea
In (\ref{phi-cor}) is definition of the color wave-function which reduces 
(\ref{color}) to
\bea
  {\cal C}_{1} & = & \frac{1}{64}f^{\mu\sigma\lambda}
  f^{\nu\rho\lambda}\,\, \delta^{\mu\tau}\delta^{\nu\xi}
  \delta^{\rho\xi}\delta^{\sigma\tau}\nn\\
  {\cal C}_{2} & = &\frac{1}{64} f^{\mu\sigma\lambda}
  f^{\nu\rho\lambda}\,\, \delta^{\mu\tau}\delta^{\nu\xi}
  \delta^{\rho\tau}\delta^{\sigma\xi}\nn\\
  {\cal C}_{3} & = &\frac{1}{64}f^{\mu\sigma\lambda}
  f^{\nu\rho\lambda}\,\, \delta^{\mu\nu}\delta^{\tau\xi}
  \delta^{\sigma\tau}\delta^{\rho\xi}\nn\\
  {\cal C}_{4} & = & \frac{1}{64}f^{\mu\sigma\lambda}
  f^{\nu\rho\lambda}\,\, \delta^{\mu\xi}\delta^{\nu\tau}
  \delta^{\tau\xi}\delta^{\rho\sigma}
\label{color2}
\eea
The quantities  $f^{abc}$ are $SU(3)$ coefficients. After total
contraction of the color indexes in (\ref{color2}) one finds
\begin{eqnarray}
  {\cal C}_{1} & = & 0 
\,\,\,\,\,\,\,\,\,\,,\,\,\,\,\,\,\,\,\,\,
  {\cal C}_{2} =  -\frac{3}{8} \nn\\
  {\cal C}_{3} & = & \frac{3}{8} 
\,\,\,\,\,\,\,\,\,\,,\,\,\,\,\,\,\,\,\,\,
  {\cal C}_{4} =  \frac{3}{8}.
  \label{aplic_41}
\end{eqnarray}
The  zero in    ${\cal C}_{1}$  of (\ref{aplic_41}) is an expected
result  as the consequence of exchanging a  color object (gluon) between
two white objects (glueballs) 

The vector $\vec{S}$  is the glueball's total spin which can be written in terms 
of the constituent gluon spin $\vec{S}_1$ e $\vec{S}_2$
\bea
  \vec{S}^{2} &=& 
 S_1^2 + S_2^2 + 2 \vec{S}_1 \cdot \vec{S}_2 \,\,.
  \label{spin}
\eea
Using the following quantum-mechanical property, for spin-one
particles $\vec{S}_i^{2} = 2$,
eq. (\ref{spin}) can be written as
\bea
  \vec{S}^{2}  = 4 + 2 \vec{S}_1\cdot \vec{S}_2.
  \label{spinglueball}
\eea
After the substitution of (\ref{spinglueball}) in (\ref{omega}) one obtains
\bea
  \omega_{1}(\mu\nu;\sigma\rho)&=&
  \frac{19}{12}\delta_{\mu\sigma}\delta_{\nu\rho}
+\frac{2}{3}\vec{S}_{1\,\mu\sigma}
  \cdot\vec{S}_{2\,\nu\rho}
  \label{ws1}
\\
  \omega_{2}(\mu\nu;\sigma\rho)&=& -\frac{7}{3}\delta_{\mu\sigma}\delta_{\nu\rho}
-
  \frac{5}{3}\vec{S}_{1\,\mu\sigma} \cdot \vec{S}_{2\,\nu\rho}.
  \label{ws2}
\eea
The spin contribution for each diagram is 
\bea
\omega_{1}^{(1)}&=& \omega_{1}(\mu\nu;\sigma\rho)\,
             \chi^{\star\mu\tau}_{A_\alpha}
         \chi^{\star\nu\xi}_{A_\gamma}
             \chi^{\rho\xi}_{A_\delta}
         \chi^{\sigma\tau}_{A_\beta} \nn\\
\omega_{1}^{(2)}&=& \omega_{1}(\mu\nu;\sigma\rho)\,
             \chi^{\star\mu\tau}_{A_\alpha}
         \chi^{\star\nu\xi}_{A_\gamma}
             \chi^{\rho\tau}_{A_\delta}
         \chi^{\sigma\xi}_{A_\beta} \nn\\
\omega_{1}^{(3)}&=& \omega_{1}(\mu\nu;\sigma\rho)\,
             \chi^{\star\mu\nu}_{A_\alpha}
         \chi^{\star\tau\xi}_{A_\gamma}
             \chi^{\sigma\tau}_{A_\delta}
         \chi^{\rho\xi}_{A_\beta} \nn\\
\omega_{1}^{(4)}&=& \omega_{1}(\mu\nu;\sigma\rho)\,
             \chi^{\star\mu\xi}_{A_\alpha}
         \chi^{\star\nu\tau}_{A_\gamma}
             \chi^{\tau\xi}_{A_\delta}
         \chi^{\rho\sigma}_{A_\beta}
\label{aplic_42}
\eea
and
\bea
\omega_{2}^{(1)}&=& \omega_{2}(\mu\nu;\sigma\rho)\,
             \chi^{\star\mu\tau}_{A_\alpha}
         \chi^{\star\nu\xi}_{A_\gamma}
             \chi^{\rho\xi}_{A_\delta}
         \chi^{\sigma\tau}_{A_\beta} \nn\\
\omega_{2}^{(2)}&=& \omega_{2}(\mu\nu;\sigma\rho)\,
             \chi^{\star\mu\tau}_{A_\alpha}
         \chi^{\star\nu\xi}_{A_\gamma}
             \chi^{\rho\tau}_{A_\delta}
         \chi^{\sigma\xi}_{A_\beta} \nn\\
\omega_{2}^{(3)}&=& \omega_{2}(\mu\nu;\sigma\rho)\,
             \chi^{\star\mu\nu}_{A_\alpha}
         \chi^{\star\tau\xi}_{A_\gamma}
             \chi^{\sigma\tau}_{A_\delta}
         \chi^{\rho\xi}_{A_\beta} \nn\\
\omega_{2}^{(4)}&=& \omega_{2}(\mu\nu;\sigma\rho)\,
             \chi^{\star\mu\xi}_{A_\alpha}
         \chi^{\star\nu\tau}_{A_\gamma}
             \chi^{\tau\xi}_{A_\delta}
         \chi^{\rho\sigma}_{A_\beta}.
\label{aplic_42a}
\eea
The glueball's spin wave-function can be written in the following  notation 
\bea
|S,S^{3}\rangle_{\rm G}
=\chi^{s_{1}s_{2}}_{S, S^{3}  }|s_1,s_2\rangle_{\rm g} \,\,. 
\label{s-state1}
\eea
The coupling of two spin-one particles  $s_1$ and $s_2$  in a total angular
momentum state of $\vec{S}=0$ or  $\vec{S}=2$ can be written using 
 angular momentum addition rules. In this sense for a state with
$S=0$ and $S^3=0$ one has 
\bea
|0,0\rangle_{\rm G}=\frac{1}{\sqrt{3}}\left[ \,\frac{}{} 
|1,-1\rangle_{\rm g} -|0,0\rangle_{\rm g}+|-1,1\rangle_{\rm g}
\,\right].
\label{s-state2}
\eea
By comparison of state (\ref{s-state1}) with (\ref{s-state2}) one finds that
\bea
\chi^{1,-1}_{0,0}=\frac{1}{\sqrt{3}}
\,\,\,\,\,\,,\,\,\,\,\,\,
\chi^{0,0}_{0,0}=-\frac{1}{\sqrt{3}}
\,\,\,\,\,\,,\,\,\,\,\,\
\chi^{-1,1}_{0,0}=\frac{1}{\sqrt{3}}.
\label{s-state3}
\eea
The state with $S=2$ has five projections $S^3=$-2,-1,0,1,2 which yields
\bea
|2,2\rangle_{\rm G}&=&|1,1\rangle_{\rm g}\nn\\
|2,1\rangle_{\rm G}&=&\frac{1}{\sqrt{2}}\left[\frac{}{} \, |1,0\rangle_{\rm g} 
+|0,1\rangle_{\rm g} \right] \nn\\
|2,0\rangle_{\rm G} &=& \frac{1}{ \sqrt{6} } |1,-1\rangle_{\rm g} +
\sqrt{ \frac{3}{2}  }|\,0,0\rangle_{\rm g}
+ \frac{1}{\sqrt{6}}  |-1,1\rangle_{\rm g} \nn\\
|2,-1\rangle_{\rm G}&=&\frac{1}{\sqrt{2}}\left[ \,
\frac{}{} |1,0\rangle_{\rm g} +|0,1\rangle_{\rm g} \right]   \nn \\
|2,-2\rangle_{\rm G}&=&|-1,-1\rangle_{\rm g} \,.
\label{s-state4}
\eea
By comparison of state (\ref{s-state1}) with (\ref{s-state4}) one finds that
\bea
\chi^{1,1}_{2,2}&=&1   \nn\\
\chi^{1,0}_{2,0}&=&\frac{1}{\sqrt{2}}
\,\,\,\,\,\,,\,\,\,\,\,\
\chi^{0,1}_{2,1}=\frac{1}{\sqrt{2}} \nn\\
\chi^{1,-1}_{2,0}&=&\frac{1}{\sqrt{6}}
\,\,\,\,\,\,,\,\,\,\,\,\,
\chi^{0,0}_{2,0}=\sqrt{\frac{3}{2}}
\,\,\,\,\,\,,\,\,\,\,\,\
\chi^{-1,1}_{2,0}=\frac{1}{\sqrt{6}}\nn\\
\chi^{0,-1}_{2,-1}&=&\frac{1}{\sqrt{2}}
\,\,\,\,\,\,,\,\,\,\,\,\
\chi^{-1,0}_{2,-1}=\frac{1}{\sqrt{2}} \nn\\
\chi^{-1,-1}_{2,-2}&=&1 . 
\label{s-state5}
\eea
Using these definitions, after contraction, of the spin indexes
one finds that for $S=0$ 
\bea
  \omega_1^{(1)} &=& 1.58333326 \nn\\
  \omega_1^{(2)} &=& 0.972222179 \nn\\
  \omega_1^{(3)} &=& 0.675925895 \nn\\
  \omega_1^{(4)} &=& 0.675925895
  \label{w10pp}
\eea
\bea
  \omega_2^{(1)} &=& -2.33333309 \nn\\
  \omega_2^{(2)} &=& -1.8888887 \nn\\
  \omega_2^{(3)} &=& -1.14814803 \nn\\
  \omega_2^{(4)} &=& -1.14814803\,.
  \label{w20pp}
\eea
For $S=2$, the present calculation considered  the unpolarized cross-section which implies
in performing an average over the spin states  resulting in
\bea
  \omega_1^{(1)} &=& 1.58333334 \nn\\
  \omega_1^{(2)} &=& 1.46228344 \nn\\
  \omega_1^{(3)} &=& 1.18539605 \nn\\
  \omega_1^{(4)} &=& 1.18539605
  \label{w12pp}
\eea
\bea
  \omega_2^{(1)} &=& -2.33333321 \nn\\
  \omega_2^{(2)} &=& -2.24529693 \nn\\
  \omega_2^{(3)} &=& -1.55307847 \nn\\
  \omega_2^{(4)} &=& -1.55307847\,.
  \label{w22pp}
\eea

The evaluation of the spatial part of the amplitude $h_{fi}$ is performed in the momentum
space, where the scattering is described in the   center of mass with 
the  CM variables: $(\vec{p},\vec{p}^{\, '})$. The OGEP has the following property
\bea
V_{\aa}(\vec{p})=V_{\aa}(-\vec{p}).
\label{par}
\eea
In this form the spatial contribution for each diagram of  figure \ref{hfi_fig} is obtained
\bea
&&h_{1}(\vec{p},\vec{p}^{\,'},\omega_{1}^{(1)},\omega_{2}^{(1)} ) =
4\,V_{\aa}  (\vec{p}-\vec{p}^{\,\prime}) 
\exp\left[-\frac{(\vec{p}-\vec{p}^{\,\prime})^{2}}{8b^{2}}\right]\nn\\
&& +4 V_{\aa}  (\vec{p}+\vec{p}')
  \exp\left[-\frac{(\vec{p}+\vec{p}')^{2}} {8b^{2}}\right]
\nn\\
&& h_{2}(\vec{p},\vec{p}^{\,'},\omega_{1}^{(2)},\omega_{2}^{(2)} ) = \frac{4
  }{(2\pi)^{3/2}b^{3}} \exp \left[-\frac{(p^{2}
  +p'^{2})}{4b^{2}}\right] 
\nn\\
&&\times\int d\vec{q}\,V_{\aa}(\vec{q})
\left\{
 \exp  \left[-\frac{q^{2}}{2b^{2}} -\frac{q\cdot(\vec{p}-\vec{p}^{\,\prime}  )}{2b^{2}}\right]
\right.\nn\\
&&\left.
+ \exp  \left[-\frac{q^{2}}{2b^{2}} -\frac{q\cdot(\vec{p}+\vec{p}^{\,\prime} )}{2b^{2}}\right]
\right\}\nn\\
&& h_{3}(\vec{p},\vec{p}^{\,'},\omega_{1}^{(3)},\omega_{2}^{(3)} ) =
\frac{4}{(2\pi)^{3/2}b^{3}}
  \exp\left[-\frac{(p^{2} +p'^{2})}{4b^{2}}\right]
\nn\\
&&\times\int d\vec{q}\,V_{\aa} (\vec{q}) 
\left\{
\exp\left[-\frac{3q^{2}}{8b^{2}}  +\frac{\vec{q}\cdot\vec{p}^{\,\prime}}{2b^{2}}\right]
\right.
\nn\\
&& +
\left.
\exp\left[-\frac{3q^{2}}{8b^{2}}  +\frac{\vec{q}\cdot\vec{p}}{2b^{2}}\right]
\right\}
\nn\\
& &h_{4}(\vec{p},\vec{p}^{\,'},\omega_{1}^{(4)},\omega_{2}^{(4)} ) =
\frac{4}{(2\pi)^{3/2}b^{3}}
  \exp\left[-\frac{(p^{2} +p'^{2})}{4b^{2}}\right]
\nn\\
&&\times\int d\vec{q}\,V_{\aa} (\vec{q}) 
\left\{
\exp\left[-\frac{3q^{2}}{8b^{2}}  +\frac{\vec{q}\cdot\vec{p}^{\,\prime } }{2b^{2}}\right]
\right.\nn\\
&& \left.
+
\exp\left[-\frac{3q^{2}}{8b^{2}}  +\frac{\vec{q}\cdot\vec{p}}{2b^{2}}\right]
\right\}\,.
\eea
The final form for the scattering amplitude is written as
\bea
h_{fi}(\vec{p},\vec{p}^{\,'})= 
\sum_{i=1}^{4} {\cal C}_{i}\,h_{i}(\vec{p},\vec{p}^{\,'},\omega_{1}^{(i)},\omega_{2}^{(i)} )
\,,
\label{hfi_sum}
\eea
from space, color and spin calculation one finds
\bea
{\cal C}_1 &=&0 \,\,\,\,\,\,\,\,\,,\,\,\,\,\,\,\,\,\, 
{\cal C}_3 = {\cal C}_4
\nn\\
\omega_{i}^{(3)}&=&\omega_{i}^{(4)} \,\,\,\,\,\,\,\,\,,\,\,\,\,\,\,\,\,\,
h_{3}=h_{4}
\eea
 which results
\bea
h_{fi}(\vec{p},\vec{p}^{\,\prime})=
 {\cal C}_{2}\,h_{2}(\vec{p},\vec{p}^{\,\prime},\omega_{1}^{(2)},\omega_{2}^{(2)} )
+2 \,{\cal C}_{3}\,h_{3}(\vec{p},\vec{p}^{\,\prime},\omega_{1}^{(3)},\omega_{2}^{(3)} )\,,
\label{hfi-fim}
\eea
or written in terms of the Mandelstam variables $(s,t)$
\bea
h_{fi}(s,t)=
 {\cal C}_{2}\,h_{2}(s,t,\omega_{1}^{(2)},\omega_{2}^{(2)} )
+2 \,{\cal C}_{3}\,h_{3}(s,t,\omega_{1}^{(3)},\omega_{2}^{(3)} )\,.
\label{hfi-fim-st}
\eea


\section*{References}
\begin{thebibliography}{}
\bibitem{Fritsch75} Fritsch H, Minkowski P 1975 {\it Nuov. Cim.} {\bf 30A} 393.
\bibitem{Vento05}  Vento V 2005 arXiv:nucl-th/0509102.
 
\bibitem{amsler1} Amsler C and Close F E 1995 {\it Phys. Lett.} B{\bf 353} 385.
\bibitem{amsler2} Amsler C and Close F E 1996 {\it Phys. Rev.} D{\bf 53} 295.
\bibitem{uthoma} Thoma U 2003 {\it Eur. Phys. J.}  A{\bf 18} 135.
\bibitem{barberis} Barberis D. et al. 2000 {\it Phys. Lett.} B{\bf 479} 345.
\bibitem{mstar}  Morningstar C J and Peardon M, 1999 {\it Phys.  Rev.} D{\bf 60} 034509.
\bibitem{bali} Bali G et al. (UKQCD) 1993{\it Phys. Lett.} B{\bf 309} 378.
\bibitem{bali2} \dash   2000  {\it Phys. Rev.} D{\bf 62} 054503.

\bibitem{cs1} Cornwall J M  and  Soni A 1983  {\it Phys. Lett.} B{\bf 120} 431.
\bibitem{cs2} Hou W S and Soni A 1984  {\it Phys. Rev.}  D{\bf 29} 101.
\bibitem{cs3} Hou W S, Luo C S and Wong G G 2001  {\it Phys. Rev.}  D{\bf 64} 014028.
\bibitem{cs4} Hou W S  and Wong G G 2003  {\it Phys. Rev.}  D{\bf 67} 034003.

\bibitem{belle} Huang H G 2001 hep-ex/0104024
\bibitem{bes} Bai J Z {\it et al} 2003 hep/ex/0307058
\bibitem{close} Close F E and Kirk A 2001  {\it Eur. Phys. J.}  C{\bf
  21} 531.
\bibitem{annals}Hadjimichef D, Krein G, Szpigel S  and da Veiga J S 1998
{\it Ann. of Phys.} {\bf 268} 105.
\bibitem{sergio} Szpigel S 1995
  {\it Intera\c{c}\~ao M\'eson-M\'eson no Formalismo Fock-Tani.}  PhD thesis 
(Doutorado em Ci\^encias) - Instituto de F\'{\i}sica, Universidade de 
S\~ao Paulo, S\~ao Paulo.
\bibitem{SilvaHB06} da Silva M L L , Hadjimichef D,  Bodmann B E J  2005 {\it article in preparation}.
\bibitem{oka1} Oka M and Yazaki K  1981 {\it Prog. Theor. Phys.} {\bf 66} 556.
\bibitem{oka2} Oka M and Yazaki K  1981 {\it Prog. Theor. Phys.} {\bf 66} 572.
\bibitem{QBD1} Swanson E S  1992  {\it Ann. of Phys.}  {\bf 220} 73.
\bibitem{QBD2} Barnes T  and Swanson E S  1992 {\it Phys. Rev.}   D{\bf 46} 131.
\bibitem{QBD3} Barnes T, Capstick S, Kovarik M D and Swanson E S 1993 {\it Phys. Rev. }
C{\bf 48} 539.
\bibitem{dimi2}   Hadjimichef D, Krein G, Szpigel S  and da Veiga J S   1996 {\it Phys.
Lett.} B{\bf 367} 317.
\bibitem{goldman1} Wu G H, Teng L J, Ping J L, Wang F, Goldman T,
 1996 {\it Phys. Rev.}  C{\bf 53} 1161.
\bibitem{goldman2} Ping J L, Wang F, Goldman T 1999  {\it Nucl. Phys.} A{\bf 657} 95.
\bibitem{dimi6}  Hadjimichef D, Haidenbauer J, Krein G 2001 {\it Phys. Rev.}  C{\bf 63}
035204; e-Print Archive: nucl-th/0010044.
\bibitem{dimi7} Hadjimichef D, Haidenbauer J, Krein G  2002 {\it Phys. Rev.} C{\bf 66}
055214.
\bibitem{dimi9}   da Silva D T , Hadjimichef D, 2004 {\it J. Phys.} G{\bf 30} 191.

\bibitem{Hansen98}  Hansen T H,  Wirstam J, Zahed I, 1998 {\it Phys. Rev.} D{\bf 58} 065012 .

\end{thebibliography}
\end{document}